\documentclass[journal]{IEEEtran}
\usepackage{amsmath}
\usepackage{graphicx}
\usepackage{subfigure}
\usepackage{graphicx}
\usepackage{epstopdf}
\usepackage{verbatim}
\usepackage{color}
\usepackage{stfloats}

\ifCLASSINFOpdf

\fi

\begin{document}

\title{A Novel RIS-Assisted Modulation Scheme}

\author{Liang Yang, Fanxu Meng, Mazen O. Hasna, and Ertugrul Basar

\thanks{L. Yang and F. Meng are with the College of Computer Science
and Electronic Engineering, Hunan University, Changsha 410082, China,
(e-mail:liangy@hnu.edu.cn, mengfx@hnu.edu.cn).}
\thanks{M. O. Hasna is with the Department of Electrical Engineering,
Qatar University, Doha 2713, Qatar. (e-mail: hasna@qu.edu.qa).}
\thanks{E. Basar is with the CoreLab, Department of Electrical and Electronics Engineering, Koç University, Istanbul 34450, Turkey (e-mail: ebasar@ku.edu.tr).}}

\maketitle

\begin{abstract}
In this work, in order to achieve higher spectrum efficiency,
we propose a reconfigurable intelligent surface (RIS)-assisted multi-user communication uplink system.
Different from previous work in which the RIS only optimizes the
phase of the incident users's signal, we propose the use of the RIS to create a
virtual constellation diagram to transmit the data of an additional user signal.
We focus on the two-user case and develop a tight approximation for the
cumulative distribution function (CDF) of the received signal-to-noise ratio
of both users. Then, based on the proposed statistical distribution, we derive
the analytical expressions of the average bit error rate of the considered two users.
The paper shows the trade off between the performance of the two users against each
other as a function of the proposed phase shift at the RIS.
\end{abstract}

\begin{IEEEkeywords}
RIS, Average BER, Spectrum Efficiency.
\end{IEEEkeywords}

\IEEEpeerreviewmaketitle

\section{Introduction}
Reconfigurable intelligent surfaces (RISs) are man-made surfaces
composed of electromagnetic (EM) materials, which are highly controllable
by leveraging electronic devices. In essence, an RIS can deliberately control
the reflection/scattering characteristics of the incident wave to
enhance the signal quality at the receiver, and hence converts the
propagation environment into a smart one [1].

Owing to their promising gains, recently RISs have been extensively
investigated in the literature. In particular, the authors in [2] proposed a practical phase
shift model for RISs. In [3], the authors studied the beamforming
optimization of RIS-assisted wireless communication under the constraints of
discrete phase shifts, while in [4], the authors studied the coverage and
signal-to-noise ratio (SNR) gain of RIS-assisted communication systems.
In [5], the authors proposed highly accurate closed-form approximations to channel
distributions of two different RIS-based wireless system setups.
Recently, RISs have been used in many scenarios and have shown
superior performance over systems not employing RISs. For instance, in [6],
an intelligent
reflecting surface (IRS)-assisted multiple-input single-output communication
system is considered, and in [7], the physical layer security of
RIS-assisted communication with an eavesdropping user is studied.
In [8], the authors proposed RIS-assisted dual-hop unmanned aerial vehicle (UAV) communication
systems while in [9], the authors used the RIS for downlink multi-user
communication from a multi-antenna base station, and developed an
energy-efficient designs for transmit power allocation and phase shifts of
the surface reflection elements.

In all of the above studies, the advantages of RISs are mainly used to
enhance the quality of the signal, and the reflection patterns were not used
to carry additional information. i.e., the role of an RIS has been mainly based
on the mitigation of the phase shifts of the involved channels, without any
additional purpose of controlling those phase shifts. In this paper,
we propose a novel modulation scheme utilizing the phase shifts of the
RIS in a spectrally efficient way to superimpose the data of an additional
user 2 (U2) on that of the ordinary user 1 (U1).

More specifically, we consider a multi-user uplink
scenario and mainly consider the feasibility of uploading the data of two users simultaneously
through the RIS, where U1 sends the data to the base
station through a direct link and is given a chance to utilize an available
RIS-assisted link to enhance its signal, but with the condition of having the data of U2
embedded with its data through the RIS. Basically, the RIS optimizes the phase of the incident
U1's signal to mitigate the phase shifts of the cascaded link as well as its direct link,
and additionally to embed the data of U2 through creating a modified virtual constellation diagram.
Hence, we assume that U2's data is known when the RIS optimizes the phase of the incident
U1's signal, and then, a virtual constellation diagram is created by the RIS
to embed U2's data. Consequently, the signal reflected by the RIS contains the data of both users.
In summary, the main contributions
of this work include the following: (i) we propose a novel and spectrallly efficient RIS-assisted modulation scheme,
(ii) for the proposed system, we develop two tight approximate statistical distributions for the
received SNR of the two considered users, (iii) based on the
proposed statistical distributions, closed-form expressions for the average
bit error rates (BER) are derived and analysed.

The remaining of this letter is organized as follows. Section II presents the system and channel models. The performance analysis is presented in section III, and the numerical and simulation results are detailed in section IV. Finally, conclusions are drawn in section V.

\section{System and Channel Models}

We consider the uplink system shown in Fig. 1, where U1 is communicating with the
base station (BS) directly and with the help of an RIS to boost its connectivity.
In the same time, U2 is in the vicinity of the RIS and is communicating with the
BS through superimposing its signal on that of U1 using the RIS.
It is assumed that the RIS can obtain perfect channel state information (CSI) through a
control link that enables it to optimize the phase shifts of the reflected signals. As
will be discussed later, the phase shifts will be utilized to superimpose U2 data on
that of U1 in a way to efficiently utilize the same spectrum. This is in return of
allowing U1 to take advantage of the RIS to improve its connectivity to the BS.
At the receiving end, the BS first decodes the signal of U1 and extracts it from the composite
received signal. In the second step, the remaining signal is processed to get the data of U2.

\begin{figure}[t]
\centering
\includegraphics[width=8cm,height=4cm]{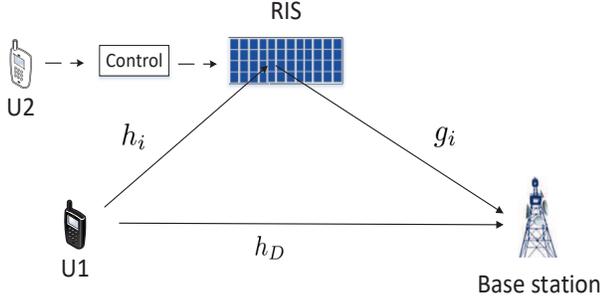}
\caption{RIS-assisted multi-user uplink.}
\label{1}
\end{figure}
\subsection{Analysis of User 1 SNR}
As mentioned above, U1 is communicating with the BS through the RIS-assisted dual-hop link
and a direct link. A binary phase shift keying (BPSK) symbol $x$ with average power $E_s$ is sent to the RIS with $N$ reflecting elements through a set of channels $h_{i} = \alpha _{i}e^{j\theta_{i}}$ where $\alpha_{i}$'s are independent and identically
distributed (i.i.d.) Rayleigh random variables (RVs) with mean $\sqrt{\pi}/2$, variance $(4-\pi)/4$, and uniformly distributed phase $\theta_{i}$. Meanwhile, the direct link to the BS is denoted as $h_{D} = \varepsilon e^{j\eta }$, where
$\varepsilon$ is a Rayleigh random variable (RV) with mean $\sqrt{\pi}/2$, variance $(4-\pi)/4$ and phase $\eta$. The RIS elements are assumed to have a line of sight with the BS through channels $g_{i} = \beta  _{i}e^{j\psi_{i}}$ where $\beta _{i}$ are i.i.d. Rician RVs with Rician factor $K$ and phases $\psi_i$. With the knowledge of the different channels CSI (U1-RIS, RIS-BS, U1-BS), the RIS optimizes the incident signals in a way to create a virtual
constellation diagram by embedding the signal of U2. The overall received signal at the BS including that of the direct link can be expressed as
\begin{align}
y_{1} &= \left ( \sqrt{\frac{E_{s}}{L_{1}}}h_{D} + \sqrt{\frac{E_{s}}{L_{2}}}
\sum_{i=1}^{N}  h_{i}g _{i} {\rm{e}}^{j\phi _{i}}\right )x+n \nonumber\\
&= \left ( \sqrt{\frac{E_{s}}{L_{1}}} \varepsilon + \sqrt{\frac{E_{s}}{L_{2}}}
\sum_{i=1}^{N}  \alpha_{i}\beta _{i} {\rm{e}}^{jw_{m}}\right )x+n,
\tag{1}\label{1}
\end{align}
where $L_{1}$ and $L_{2}$
are the path losses of the direct link and the RIS-assisted dual-hop link, respectively, $\phi_{i} = (w_{m}- \theta_{i} -
\psi_{i}+ \eta)$ is the adjustable phase introduced by the $i$th reflecting
element of the RIS to mitigate the channels' phase shifts, $w_{m}$ is the message-dependent phase
introduced by the RIS to carry the information of U2 where $w_{m}$
represents a binary symbol of 1 and $-w_{m}$ represents 0, and $n \sim \mathcal C
\mathcal N(0,N_{0})$ is the additive white Gaussian noise (AWGN) signal.

Then, the received SNR can be written as
\begin{align}
\gamma_{1} = &\frac{\left ( \sqrt{\frac{E_{s}}{L_{1}}}\varepsilon + \sqrt{\frac{E_{s}}{L_{2}}}
\sum_{i=1}^{N}  \alpha_{i}\beta _{i} \cos(w_{m})\right )^2}{N_{0}}\nonumber\\
=&\left (\sqrt{\bar\gamma_{1}}\varepsilon + \sqrt{\bar\gamma_{2}}R_{1}\right)^2 = Z_{1}^{2},
\tag{2}\label{2}
\end{align}
where $R_{1} =\sum_{i=1}^{N}  \alpha_{i}\beta _{i} \cos(w_{m})$ ,  $\bar{\gamma}_{1}={\bar{\gamma}}/{L_{1}}$, $\bar{\gamma}_{2}={\bar{\gamma}}/{L_{2}}$ and $\bar \gamma = E_{s}/N_{0}$ denotes the average SNR.

For a sufficiently large number of reflecting elements $N$, and relying on the
central limit theorem (CLT), $R_{1}$ can be assumed to follow a Gaussian RV. Thus, the probability density function (PDF) of $R_{1}$ can be simply expressed as

\begin{equation}
f_{R_{1}}(r) = \frac{1}{\sqrt{2\pi\sigma_{1}^2}}\exp\left
( -\frac{\left ( r-\mu _{1} \right )^2}{2\sigma_{1}^2} \right ),
\tag{3}\label{3}
\end{equation}
where $\mu _{1}=\frac{N \Gamma\left ( \frac{3}{2} \right )\sqrt{\pi}}{2\sqrt{1+K}e^{K}}{}_{1}F_{1}
(\frac{3}{2},1,K)\cos(w_{m})$, $\sigma _{1}^2 = N\left [  \frac{{}_{1}{\rm{F}}_{1}
(2,1,K)}{(1+K)
e^{K}} - \left (\frac{\mu _{1}}{N\cos(w_{m})}  \right )^2 \right ]\cos^{2}(w_{m})$,
${}_{1}F_{1}(\cdot)$ is the Degenerate hypergeometric function, and $\Gamma(\cdot)$ is the Gamma function \cite{13}.
Then, we obtain the cumulative distribution function (CDF) of $Z_{1}$ as
\begin{align}
F_{Z_{1}}(z) &= \Pr(\sqrt{\bar\gamma_{1}}\varepsilon +\sqrt{\bar\gamma_{2}}R_{1} < Z)\nonumber\\
&=\int_{0 }^{\frac{z}{\sqrt{\bar\gamma_{2}}}}\frac{1}
{\sqrt{2\pi\sigma_{1}^2}}\exp\left ( -\frac{\left ( r-\mu _{1} \right )^2}{2\sigma_{1}^2} \right )dr\nonumber\\
&-\int_{0 }^{\frac{z}{\sqrt{\bar\gamma_{2}}}}\frac{1}
{\sqrt{2\pi\sigma_{1}^2}} {\rm{e}}^{\left ( -\frac{\left ( r-\mu _{1}
\right )^2}{2\sigma_{1}^2} -\frac{\left ( z-\sqrt{\bar\gamma_{2}}r \right )^2}{\bar\gamma_{1}}\right )}dr.
\tag{4}\label{4}
\end{align}

From [10, Eq.(2.33.1)], the CDF of $\gamma_1$ is given by
\begin{align}
&F_{\gamma_{1}}(\gamma ) = \frac{1}{2}\left [ \rm{erf}\left
( \frac{C_{2}\sqrt{\gamma}-\mu_{1}}{\sqrt{2\sigma_{1}^2}} \right ) +
\rm{erf}\left ( \frac{\mu_{1}}{\sqrt{2\sigma_{1}^2}} \right )\right ]\nonumber\\
&-\frac{1}{2\sqrt{2}C_{3}}\rm{e}^{\left ( -\frac{\left (C_{1} C_{2}\sqrt{\gamma}-
C_{1}\mu_{1} \right )^{2}}{2C_{3}^{2}C_{2}^{2}} \right )}\left [ \rm{erf}\left (
\frac{C_{2}\sqrt{\gamma}-\mu_{1}}{2C_{3}\sigma_{1}} \right ) \right ]\nonumber\\
&- \frac{1}{2\sqrt{2}C_{3}}\rm{e}^{\left ( -\frac{\left (C_{1} C_{2}\sqrt{\gamma}-
C_{1}\mu_{1} \right )^{2}}{2C_{3}^{2}C_{2}^{2}} \right )}\left [ \rm{erf}\left
( \frac{C_{4}\sqrt{\gamma}+\mu_{1}}{2C_{3}\sigma_{1}} \right ) \right ],
\tag{5}\label{5}
\end{align}
where $C_{1}=\frac{1}{\sqrt{\bar\gamma_{1}}}$,
$C_{2}=\frac{1}{\sqrt{\bar\gamma_{2}}}$,$C_{3} = \sqrt{\frac{\sigma^2L_{1}}
{L_{2}}+\frac{1}{2}}$, $C_{4} = \frac{2\sigma_{1}^2 C_{1}^2}{C_{2}}$ and
${\rm{erf}}(\cdot)$ is the error function \cite{13}.

\newcounter{TempEqCnt}
\setcounter{TempEqCnt}{\value{equation}}
\setcounter{equation}{13}
\begin{figure*}[hb]
\hrulefill
\begin{align}
A_{2}{=}\sqrt{\frac{\bar{\gamma }_{1}\gamma }{2\sigma_{2}^2\bar{\gamma }_{2}{+}
\bar{\gamma }_{1}\gamma }}\exp\left ( \frac{1}{\bar{\gamma }_{1}} {-}
\frac{\mu _{2}^2}{2\sigma_{2}^2} \right )\left[ \sum_{l=0}^{L}
\sum_{k=0}^{l-1}\frac{(\gamma \mu_{2}^2)^l \left ( \frac{1}{2}{-}l \right )_{l{-}k{-}1}
({-}1)^{l{+}k{-}1}\left ( \frac{1}{\bar\gamma_{1}}{+}
\frac{\gamma}{2\bar\gamma_{2}\sigma_{2}^2} \right )^{k{+}\frac{1}{2}} }
{l!2^l\Gamma(l{+}\frac{1}{2})\left (\gamma \sigma_{2}^2 {+}
\frac{2\bar\gamma_{2}\sigma_{2}^4}{\bar\gamma_{1}} \right )^l \exp\left (\frac{1}{\bar\gamma_{1}}{+}
\frac{\gamma}{2\bar\gamma_{2}\sigma_{2}^2} \right )}{+}\sum_{l=0}^{L}
\frac{(\gamma \mu_{2}^2)^l {\rm{erfc}}\left
( \frac{1}{\bar\gamma_{1}}{+}
\frac{\gamma}{2\bar\gamma_{2}\sigma_{2}^2} \right )}{l!2^l
\left (\gamma \sigma_{2}^2 {+} \frac{2\bar\gamma_{2}
\sigma_{2}^4}{\bar\gamma_{1}} \right )^l } \right]
\tag{12}\label{12}
\end{align}
\end{figure*}
\setcounter{equation}{\value{TempEqCnt}}

\newcounter{mytempEqCnt}
\setcounter{mytempEqCnt}{\value{equation}}
\setcounter{equation}{12}
\begin{figure*}[hb]
\hrulefill
\begin{align}
I_{1}=&\frac{\sqrt{\pi}}{2}\left [  1+{\rm{erf}}(\frac{\mu_{1}}{\sqrt{2\sigma_{1}^{2}}})
-2{\rm{erf}}(\frac{\mu_{1}}{C_{2}}) \right ]
+\sum_{i=1}^{4}\frac{S_{i}}{2}\sqrt{\frac{2\pi\sigma_{1}^2}
{2\sigma_{1}^2+T_{i}C_{2}^2}}\exp(\frac{T_{i}^2C_{2}^2\mu _{1}^2}
{4\sigma_{1}^4+2T_{i}C_{2}^2\sigma_{1}^2} -
\frac{T_{i}\mu _{1}^2}{2\sigma_{1}^2}) \nonumber\\
&\times\left[ 2{\rm{erf}}\left ( \sqrt{\frac{\mu _{1}^2}{C_{2}^2} +\frac{T_{i}^2\mu _{1}^2}{2\sigma_{1}^2}} -
\frac{T_{i}C_{2}\mu _{1}}{\sqrt{4
\sigma_{1}^4+2T_{i}C_{2}^2\sigma_{1}^2}}\right )- {\rm{erfc}}\left ( \frac{T_{i}C_{2}\mu _{1}}{\sqrt{4
\sigma_{1}^4+2T_{i}C_{2}^2\sigma_{1}^2}}\right )\right]
\tag{16}\label{16}
\end{align}

\begin{align}
I_{2} = &\frac{1}{2}\sqrt{\frac{\pi}{2C_{3}^2+C_{1}^2}}
\exp\left ( -\frac{\mu _{1}^2C_{1}^2}{2C_{2}^2C_{3}^2+C_{1}^2C_{2}^2} \right )\left (1+\rm{erf}\left ( \frac{C_{1}^2\mu_{1}}{\sqrt{4C_{2}^2C_{3}^4
+2C_{1}^2C_{2}^2C_{3}^2}} \right )  \right )\nonumber\\
&-\sum_{i=1}^{4}\frac{S_{i}}{\sqrt{2}}
\sqrt{\frac{\pi\sigma_{1}^2}{4C_{3}^2\sigma_{1}^2 + 2C_{1}^2\sigma_{1}^2+T_{i}C_{4}^2}}
\exp\left ( C_{5}- \frac{\mu _{1}^2C_{1}^2}{2C_{2}^2C_{3}^2 }-\frac{\mu _{1}^2T_{i}}{4C_{3}^2\sigma_{1}^2 } \right )\left (1+{\rm{erf}}\left ( \sqrt{C_{5}}  \right )\right )
\tag{17}\label{17}
\end{align}

\begin{align}
I_{3} = &\frac{1}{2}\sqrt{\frac{\pi}{2C_{3}^2+C_{1}^2}}
\exp\left ( -\frac{\mu _{1}^2C_{1}^2}{2C_{2}^2C_{3}^2+C_{1}^2C_{2}^2} \right )
\left [{\rm{erfc}} \left (
\frac{\mu _{1}C_{1}^2}{\sqrt{4C_{2}^2C_{3}^4+2C_{1}^2C_{2}^2C_{3}^2}} \right )
- 2{\rm{erf}}  \left ( \sqrt{C_{7}} -
\frac{\mu _{1}C_{1}^2}{\sqrt{4C_{2}^2C_{3}^4+2C_{1}^2C_{2}^2C_{3}^2}} \right )
\right ]\nonumber\\
&+\sum_{i=1}^{4}\frac{S_{i}}{\sqrt{2}}\sqrt{\frac{\pi\sigma_{1}^2}{4C_{3}^2
\sigma_{1}^2+2C_{1}^2\sigma_{1}^2+T_{i}C_{2}^2}}
\exp\left ( C_{6}-\frac{C_{1}^2\mu_{1}^2}{2C_{2}^2C_{3}^2}-\frac{T_{i}\mu_{1}^2}
{4C_{3}^2\sigma_{1}^2} \right )\left [2{\rm{erf}}  \left ( \sqrt{C_{7}} -
\sqrt{C_{6}}\right )-{\rm{erfc}} \left (\sqrt{C_{6}}\right )  \right ]
\tag{18}\label{18}
\end{align}
\end{figure*}
\setcounter{equation}{\value{TempEqCnt}}

\subsection{Analysis of User 2 SNR}
Assuming that the signal of U1 can be successfully decoded (represented by $\hat{x}$), the received signal can now be expressed as
\begin{align}
y_{2} &= y_{1} {\rm{e}}^{j\hat{x}\pi}\nonumber\\
&=\sqrt{\frac{E_{s}}{L_{2}}}\sum_{i=1}^{N}  \alpha _{i}\beta _{i} {\rm{e}}^
{jw_{m}}x {\rm{e}}^{j\hat{x}\pi} + \sqrt{\frac{E_{s}}
{L_{1}}}\varepsilon
x {\rm{e}}^{j\hat{x}\pi} + n {\rm{e}}^{j\hat{x}\pi}\nonumber\\
&=\sqrt{\frac{E_{s}}{L_{2}}}\sum_{i=1}^{N}  \alpha _{i}\beta _{i}
{\rm{e}}^{jw_{m}} + \sqrt{\frac{E_{s}}{L_{1}}}\varepsilon + n{\rm{e}}^{j\hat{x}\pi}.
\tag{6}\label{6}
\end{align}

Thus, the signal of U2 can be regarded as a biased BPSK signal with an initial
phase of $\frac{\pi}{2}$, and an offset angle of $\frac{\pi}{2} - w_{m}$. Then, the SNR of U2 can be written as
\begin{align}
\gamma_{2} = \frac{\frac{\bar{\gamma}}{L_{2}}\left (\sum_{i=1}^{N}\alpha_{i}
\beta _{i} \cos(\frac{\pi}{2}-w_{m})\right )^2}
{\frac{\bar{\gamma}}{L_{1}}\left | \varepsilon  \right |^{2} + 1}\
=\frac{\bar{\gamma}_{2}R_{2}^2}{\bar{\gamma}_{1}\left | \varepsilon  \right |^{2} + 1}.
\tag{7}\label{7}
\end{align}
where $R_{2}=\sum_{i=1}^{N}\alpha_{i}
\beta _{i} \cos(\frac{\pi}{2}-w_{m})$.
Similar to $R_{1}$, and relying again on the CLT, $R_{2}$ is assumed to follow the Gaussian
distribution with mean $\mu _{2}$ and variance $\sigma _{2}^2$.
The PDF of $R_{2}$ can be written as
\begin{equation}
f_{R_{2}}(r) = \frac{1}{\sqrt{2\pi\sigma_{2}^2}}\exp\left
( -\frac{\left ( r-\mu _{2} \right )^2}{2\sigma_{2}^2} \right ),
\tag{8}\label{8}
\end{equation}
where $\mu _{2}=\frac{N \Gamma\left ( \frac{3}{2} \right )\sqrt{\pi}}{2\sqrt{1+K}e^{K}}{}_{1}F_{1}
(\frac{3}{2},1,K)\cos(\frac{\pi}{2}-w_{m})$, and $\sigma _{2}^2 = N\left [  \frac{{}_{1}{\rm{F}}_{1}
(2,1,K)}{(1+K)
e^{K}} - \left (\frac{\mu _{1}}{N\cos(\frac{\pi}{2}-w_{m})}  \right )^2 \right ]\cos^{2}(\frac{\pi}{2}-w_{m})$.

Let $Y = \bar{\gamma}_{1}\left | \varepsilon  \right |^{2} + 1$, then the PDF of $Y$ can be readily written as $f_{Y}(y)=\frac{1}
{\bar\gamma_{1}}e^{-\frac{y-1}{\bar\gamma_{1}}}$. Thus the CDF of $\gamma_{2}$ can be calculated as
\begin{align}
F_{\gamma_{2}}(\gamma ) =&\Pr\left ( \frac{\bar{\gamma}_{2}R_{2}^2}{Y} < \gamma \right )\nonumber\\
=& 1-Q_{\frac{1}{2}}\left ( \frac{\mu _{2} }{\sqrt{\sigma_{2}^2}} ,
\sqrt{\frac{\gamma }{\sigma_{2}^2\bar{\gamma }_{2}}}\right )\nonumber\\
&+ \sqrt{\frac{\bar{\gamma }_{1}\gamma }{2\sigma_{2}^2\bar{\gamma }
_{2}+\bar{\gamma }_{1}\gamma }}\exp\left ( \frac{1}{\bar{\gamma }
_{1}} - \frac{\mu _{2}^2\bar{\gamma }_{2}}{\bar{\gamma }_{1}\gamma
 +2\sigma_{2}^2\bar{\gamma }_{2}} \right )\nonumber\\
&\times Q_{\frac{1}{2}}\left ( \sqrt{\frac{\mu _{2}^2\gamma}{\sigma_{2}^2
\gamma+\frac{2\bar{\gamma}_{2}\sigma_{2}^4}{\bar{\gamma}_{1}}}} ,
\sqrt{\frac{2}{\bar{\gamma}_{1}}+\frac{\gamma}{\bar{\gamma}_{2}\sigma_{2}^2}}\right ),
\tag{9}\label{9}
\end{align}
where $Q_{\frac{1}{2}}(\cdot,\cdot)$ is the Marcum Q-function \cite{13}.

Using (9) to calculate the average BER is difficult, however, from [11, Eq. (16)], we have
\begin{equation}
Q_{\frac{1}{2}}\left ( a,b \right ) \approx  \sum_{l=0}^{L}\frac{a^{2l}
\Gamma(l+\frac{1}{2},\frac{b^2}{2})}{l! \Gamma(l+\frac{1}{2})2^l {\rm{e}}^{\frac{a^2}{2}}},
\tag{10}\label{10}
\end{equation}
where $\Gamma(\cdot,\cdot)$ is the incomplete gamma function [10].
Then, with the help of [12, Eq. (06.06.03.0005.01)], the CDF of $\gamma_{2}$ can be written as
\begin{align}
F_{\gamma_{2}}(\gamma )  \approx   A_{1}+A_{2},
\tag{11}\label{11}
\end{align}
where $A_{1}= 1 - \sum_{l=0}^{L}\frac{(\frac{\mu _{2}}{\sigma_{2}})^{2l}
\Gamma(l+\frac{1}{2},\frac{\gamma}{2\bar\gamma_{2}\sigma_{2}^2})}{l!
 \Gamma(l+\frac{1}{2})2^l {\rm{exp}}({\frac{\mu ^2}{2\sigma_{2}^2}})}$,
  $A_{2}$ is shown at the bottom of this page, and $\rm{erfc}(\cdot)$ is the complementary error function [10].

\setcounter{equation}{18}
\begin{figure*}[ht]
\hrulefill
\begin{align}
I_{4}=& \Gamma\left(\frac{1}{2}\right)-\sum_{l=0}^{L}\left ( \frac{\left ( \frac{\mu _{2}^2}
 {\sigma_{2}^2} \right )^{l}\sqrt{2\bar{\gamma}_{2}\sigma_{2}^2}}{2^{l-1}
 \Gamma(l+\frac{1}{2})e^{\frac{\mu _{2}^2}{2\sigma_{2}^2}}(1+2\bar{\gamma}_{2}
 \sigma_{2}^2)^{1+l}} {}_{2}{\rm{F}}_{1}\left (1,l+1;\frac{3}{2};
 \frac{2\bar{\gamma}_{2}\sigma_{2}^2}{1+2\bar{\gamma}_{2}\sigma_{2}^2}  \right )\right )
 \tag{21}\label{21}
\end{align}

\begin{align}
I_{5}=&\exp\left ( \frac{1}{\bar{\gamma }_{1}} - \frac{\mu _{2}^2}
{2\sigma_{2}^2} \right )\sum_{l=0}^{L}\sum_{i=1}^{4}\frac{S_{i}
(\frac{\mu _{2}}{\sigma_{2}})^{2l}}{2^l(1+\frac{T_{i}}{2\bar{\gamma}
_{2}\sigma_{2}^2})^{3/4}(\frac{2\bar{\gamma}_{2}\sigma_{2}^2}
{\bar{\gamma}_{1}})^{1/4}}\exp\left ( \frac{\bar{\gamma}_{2}
\sigma_{2}^2}{\bar{\gamma}_{1}} - \frac{T_{i}}{2\bar{\gamma}_{1}}\right )
{\rm{W}}_{-l-\frac{1}{4},-\frac{1}{4}}\left ( \frac{2\bar{\gamma}_{2}
\sigma_{2}^2}{\bar{\gamma}_{1}} + \frac{T_{i}}{\bar{\gamma}_{1}}\right )
\tag{22}\label{22}
\end{align}

\begin{align}
I_{6} = & \exp\left ( \frac{1}{\bar{\gamma }_{1}} - \frac{\mu _{2}^2}{2\sigma_{2}^2} \right )
\sum_{l=0}^{L}\sum_{k=1}^{l-1}\frac{\mu _{2}^{2l}
(\bar{\gamma}_{2}\sigma_{2}^2)^{k/2}\left (\frac{1}{2}-l  \right )_{l-k-1}(-1)^{k+l-1}}{\Gamma(l+\frac{1}{2})2^l
\sigma_{2}^{2l}(\bar{\gamma}_{2}\bar{\gamma}_{1}\sigma_{2}^2)^{k+\frac{1}{2}}
\sqrt{\bar{\gamma}_{1}/2}}
\exp\left ( \frac{\bar{\gamma}_{2}\sigma_{2}^2}{\bar{\gamma}_{1}} -
\frac{1}{2\bar{\gamma}_{1}}\right )\left ( \frac{2}{\bar{\gamma}_{1}}
+\frac{1}{\bar{\gamma}_{2}\bar{\gamma}_{1}\sigma_{2}^2} \right )^{-k/2-1}\nonumber\\
& \times{\rm{W}}_{\frac{k-2l}{2},\frac{-k-1}{2}}\left ( \frac{2\bar{\gamma}_{2}
\sigma_{2}^2}{\bar{\gamma}_{1}} + \frac{1}{\bar{\gamma}_{1}}\right )
\tag{23}\label{23}
\end{align}
\end{figure*}
\section{Performance Analysis}
In this section, we analyze the performance of the proposed scheme by deriving closed-form expressions for the average BER.
For different binary modulation schemes, a unified average BER expression is given by \cite{16}

\begin{equation}
P_{e} = \frac{q^p}{2\Gamma (p)}\int_{0}^{\infty }
\exp(-q\gamma )\gamma ^{p-1}F_{\gamma}(\gamma )d\gamma,
\tag{13}\label{13}
\end{equation}
where $F_{\gamma}(\gamma )$ is the CDF of $\gamma$, and the
parameters $p$ and $q$ are modulation
schemes dependent. In this work, we consider BPSK modulation, and hence we use $p = \frac{1}{2}$ and $q$ = 1.

\subsubsection{Average BER of User 1}
From (\ref{5}) and (\ref{13}), $P_{e1}$ can be formulated as
\begin{align}
P_{e1} = (I_{1}-I_{2} -I_{3})/(2\Gamma (0.5)),
\tag{14}\label{14}
\end{align}
where $I_{1}$, $I_{2}$ and $I_{3}$ are derived next.
Using the expression in (\ref{5}) to evaluate $P_{e1}$ is difficult.
Hence, we opt to utilize an alternative expression for the erf function \cite{17} as
 \begin{equation}
\begin{split}
\rm{erf}(x) \approx\left\{\begin{matrix}
  1-\sum_{i=1}^{4}S_{i} {\rm{e}}^{-T_{i}x^2}  x\geq 0 \\
 -1+\sum_{i=1}^{4}S_{i} {\rm{e}}^{-T_{i}x^2}  x<  0
\end{matrix}\right.
\end{split}
\tag{15}\label{15}
\end{equation}
where $S=[\frac{1}{8},\frac{1}{4},\frac{1}{4},\frac{1}{4}]$
and $T=[1,2,\frac{20}{3},\frac{20}{17}]$.

Then, with the help of [10, Eq. (2.33.1)], closed form expressions for  $I_{1}$, $I_{2}$,
and $I_{3}$ are shown at the bottom of the previous page, where $C_{5}=\frac{\left ( 2C_{1}^2\mu_{1}\sigma_{1}^2-T_{i}C_{2}C_{4}\mu_{1} \right )^2}{16C_{2}^2C_{3}^4\sigma_{1}^4
+8C_{1}^2C_{2}^2C_{3}^2\sigma_{1}^4+4T_{i}C_{2}^2C_{3}^2C_{4}^2
\sigma_{1}^2}$, $C_{6}=\frac{\left ( 2C_{1}^2\mu_{1}\sigma_{1}^2+T_{i}C_{2}^2\mu_{1} \right )^2}
{16C_{2}^2C_{3}^4\sigma_{1}^4+8C_{1}^2C_{2}^2C_{3}^2\sigma_{1}^4+4T_{i}C_{2}^4
C_{3}^2\sigma_{1}^2}$, and $C_{7}=\frac{\mu _{1}^2}{C_{2}^2}+
\frac{\mu _{1}^2C_{1}^2}{2C_{2}^2C_{3}^2}$.

\subsubsection{Average BER of User 2}
As the decoding of U2 signal follows that of U1, it is usually
difficult to ensure that the decoded signal is completely correct.
Therefore, after processing the received signal, we might get some
inverted information bits of U2. Then, the practical average BER of U2 can be expressed as
\begin{align}
P_{e2} = P_{e2}^{ideal}(1-P_{e1}) + P_{e1}(1-P_{e2}^{ideal}),
\tag{19}\label{19}
\end{align}
where $P_{e2}^{ideal}$ denotes U2's average BER with ideal conditions (i.e.
assuming the decoded U1's data is completely correct). From (\ref{11}) and (\ref{13}),
and using [10, Eq. (6.455)], the ideal U2's average BER can be calculated as
\begin{align}
P_{e2}^{ideal} 
=(I_{4} + I_{5} + I_{6})/(2\Gamma (0.5)),
\tag{20}\label{20}
\end{align}
where $I_{4}$, $I_{5}$ and $I_{6}$ can be shown to be given as in (\ref{21}), (\ref{22}) and (\ref{23}), where
${}_{2}F_{1}(\cdot)$ is the Gauss hypergeometric function, and  $W_{a,b}(\cdot)$
is the Whittaker hypergeometric function \cite{13}.

\begin{figure} \centering
\centering
\includegraphics[width=8cm,height=5.5cm]{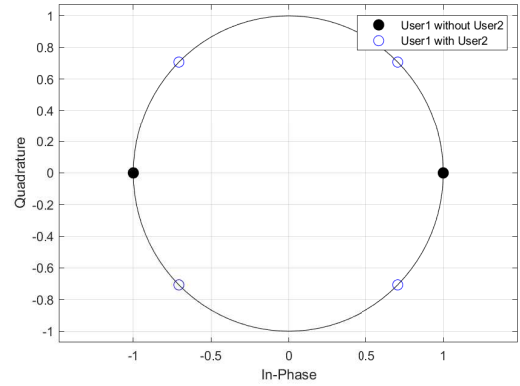}
\caption{The constellation diagram of U1 and U2.}
\label{fig:2}
\end{figure}

\begin{figure}[h]
\centering
\includegraphics[width=8cm,height=6.5cm]{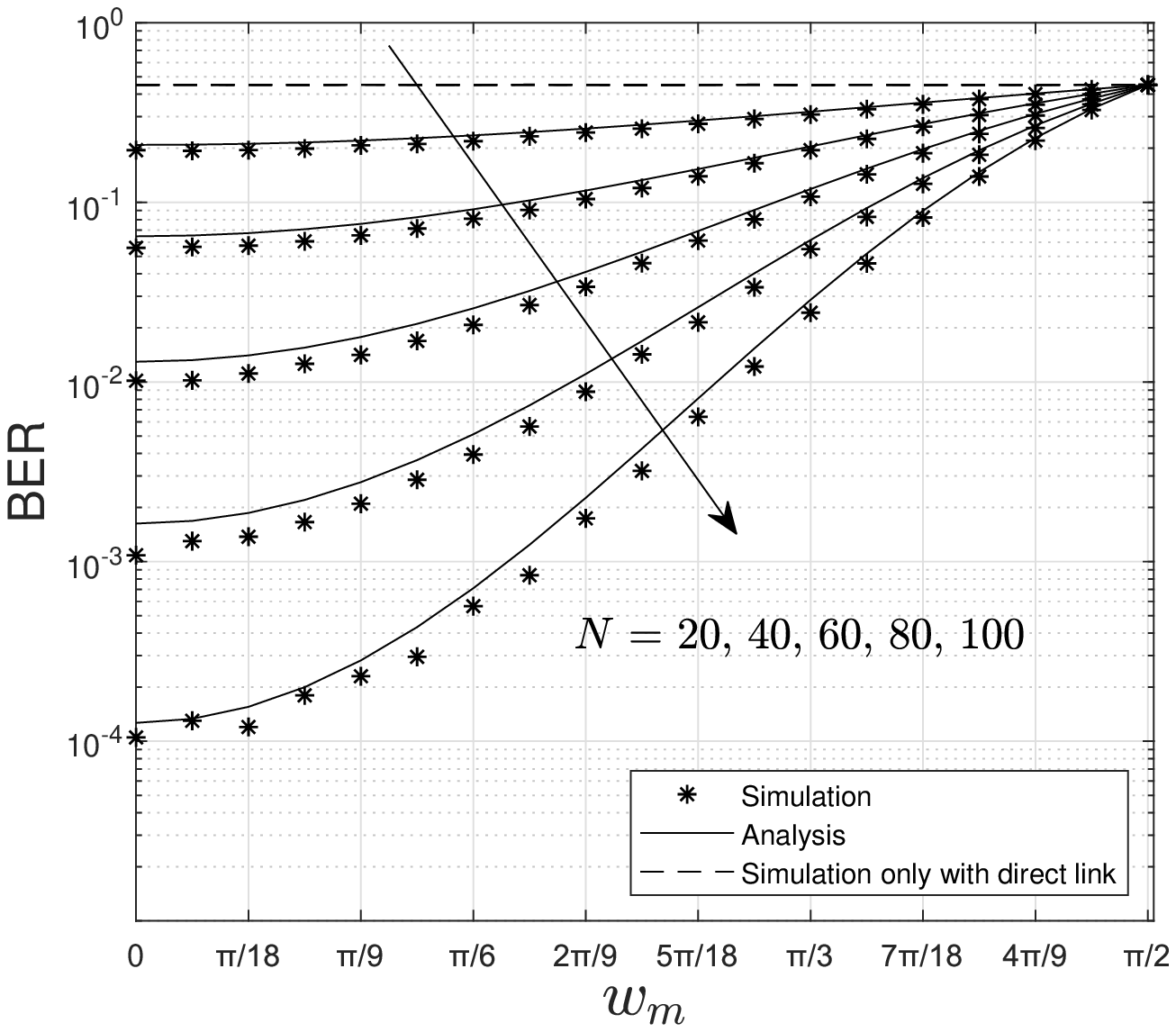}
\caption{Average BER of U1 versus $w_{m}$.}
\label{fig:3}
\end{figure}

\begin{figure}[h]
\centering
\includegraphics[width=8cm,height=6.5cm]{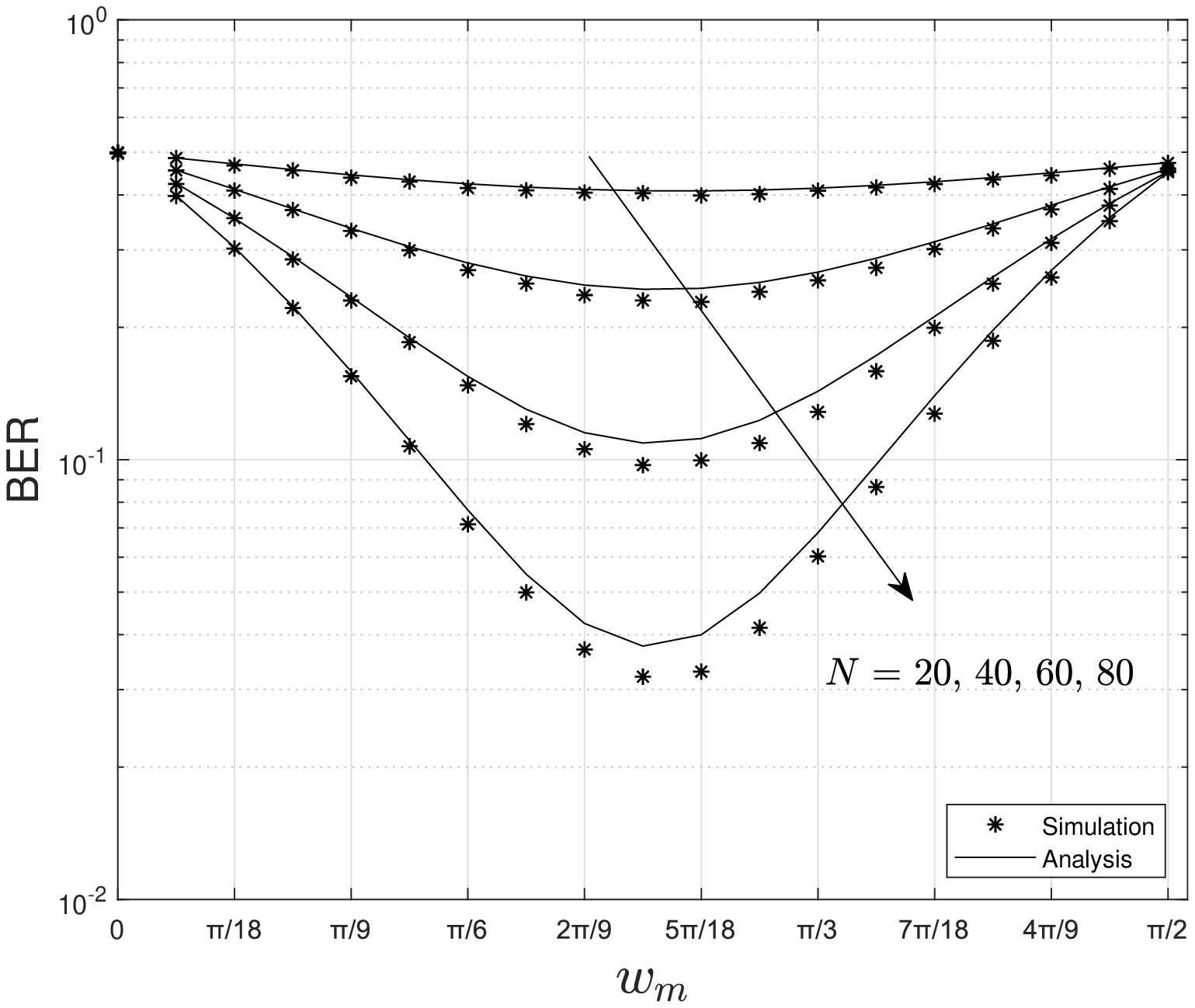}
\caption{Average BER of U2 versus $w_{m}$.}
\label{fig:4}
\end{figure}
\begin{figure}[h]
\centering
\includegraphics[width=8cm,height=6.5cm]{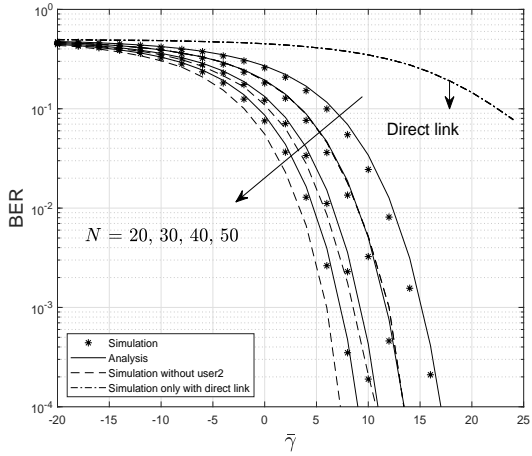}
\caption{Average BER of U1 versus $\bar\gamma$ with and without U2.}
\label{fig:5}
\end{figure}

\section{Numerical and Simulation Results}
In this section, we present some numerical results to verify
our analysis. The parameters used in the figures
are $k$ = 3, $p=\frac{1}{2}$, $q=1$, $L_{1} = 20$ dB, and $L_{3} = 30$ dB.

In Fig. \ref{fig:2} we plot the constellation diagram of U1 and U2
where $\bar\gamma = 20$ dB, $N = 50$ and $w_{m} = \pi/4$. It can be deduced from Fig. 2
that superimposing U2's data on that of U1 causes the constellation
diagram of U1's to shift based on the value of $w_{m}$ used. This
causes the BER of U1 to increase, simply because the separation of the
two constellation points is lower at this time. From the constellation of
U2, it can be deduced that the processed signal is similar to a BPSK signal where the initial phase is $-\pi/2$.

In Fig. \ref{fig:3}, we plot the average BER of U1 as a function of $w_{m}$.
When $w_{m} = 0$, this is equivalent to the case where there is no U2 and
only the data of U1 is transmitted in the
uplink, and its constellation is a pure BPSK one. As we increase $w_{m}$, the system needs to utilize the spectrum
resources of U1 to transmit the signal of U2. Consequently, U1's average BER increases. When $w_{m}$ reaches $ \pi/2$, it leads
to the lowest data accuracy of U1. In addition, and as expected, it can be seen from Fig. \ref{fig:3} that increasing
$N$ can bring performance improvements to U1. This means that we can reduce the negative effect of $w_{m}$ by increasing $N$.

In Fig. \ref{fig:4}, we plot the average BER of U2 versus $w_{m}$. It is
clear from the figure that by increasing $w_{m}$, the average BER of
U2 decreases first and then increases. When $w_{m}$ is small, $P_{e1}$
is small, and hence the system performance is mainly determined by $P_{e2}^{ideal}$. Similar to
the observations in Fig. 3, increasing $w_{m}$ causes $P_{e1}$ to increase, and
when $w_{m}$ is large, one cannot get a clean U2 signal due to the large
average BER of U1. Hence, the system performance is dominated in this
case by $P_{e1}$. This means that the choice of $w_{m}$ is critical to ensure reasonable performance of both users.

In Fig. \ref{fig:5}, we plot the average BER of U1 versus $\bar\gamma$ with and without U2.
As expected, increasing the average SNR leads to decreasing the average BER.
In addition, the effect of having U2 on the average BER of U1 is clear
from the figure. The figure shows also the performance of U1 when using the direct link only. It is clear that the incentive given to U1 through the RIS usage to allow the superposition of U2 data is worth it to U1 from performance point of view. Finally, we observe a close match between the derived expressions and the simulation results which confirms the accuracy of the analytical expressions.


\section{Conclusion}
In this letter, we proposed an RIS-assisted multi-user uplink communication system employing a novel modulation scheme.
More specifically, we derived the analytical expression of average BER and tight approximation
on the CDF of the received SNR of the case of two users sharing the same spectrum with the help of the RIS.
Numerical results show that we can obtain U2's data
with higher accuracy while ensuring the accuracy of U1's data by setting an
appropriate phase shift and large enough number of surface elements.

\end{document}